\documentclass{aa}
\usepackage{txfonts}
\usepackage{epsfig}
\usepackage{natbib}
\bibpunct{(}{)}{;}{a}{}{,}

\begin{document}

\headnote{Research Note}

\title{A three-dimensional Galactic extinction model}
\author{R. Drimmel\inst{1} \and A. Cabrera-Lavers\inst{2} \and M. L\'opez-Corredoira\inst{3}} 

\offprints{drimmel@to.astro.it} 

\institute{Istituto Nazionale di Astrofisica (INAF), Osservatorio
Astronomico di Torino, I-10025 Pino Torinese, Italy
\and Instituto de Astrofisica de Canarias, E-38200 La Laguna,
Tenerife, Spain
\and Astronomisches Institut der Universit\"at Basel, Venusstrasse 7,
CH-4102 Binningen, Switzerland}

\date{Received / Accepted}


\abstract{ 
A large-scale three-dimensional model of Galactic extinction 
is presented based on
the Galactic dust distribution model of \citet{DS01}. The
extinction $A_{\rm V}$ to any point within the Galactic disk can be quickly
deduced using a set of three-dimensional cartesian grids. Extinctions
from the model are compared to empirical extinction measures, including 
lines-of-sight in and near
the Galactic plane using optical and NIR extinction
measures; in particular we show how extinction can be derived from NIR
color-magnitude diagrams in the Galactic plane to a distance of 8 kiloparsec.
}

\maketitle

\section{Introduction}

In the past a fundamental obstacle to Galactic studies has been
extinction due to interstellar dust, which has limited our view of
the Galactic stellar distribution to the solar neighborhood. 
During the past decade NIR surveys, together with ever deeper optical
surveys, have been piercing the interstellar haze, probing the stellar
distribution in the Galactic plane and revealing its nonaxisymmetric
structure,
including the barred nature of the Galactic bulge
\citep{Blitz91,Weil94}, the Galactic warp \citep{DJO89, FRE94} and the
spiral arms \citep{Dri00}.
Newly available high-resolution NIR
surveys, such as 2MASS and DENIS, promise to further reveal the nature
of these structures. However, while the effects of
interstellar dust are mitigated at NIR wavelengths, they are 
still important.

The most recent map of Galactic extinction over the entire sky is that
of Schlegel, Finkbeiner and Davis (1998, hereafter
SFD)\nocite{Schleg98}, however it only maps the {\em total} Galactic 
extinction and is therefore most appropriate for extragalactic studies.
To study stellar populations or objects found throughout the
Galactic disk a volumetric description of Galactic extinction is
required. Three-dimensional extinction models have been produced using
reddening data from stellar samples
\citep{NK80,Arenou92,Hakkila97,Mendez98}, but these give only a local
description of extinction, being reliable to heliocentric distances of
about 2 kiloparsecs due to
limited sampling at optical wavelengths. However, in the NIR a
well-defined stellar population can provide extinction measures along a
line-of-sight to much greater distances; here we describe in detail
one such method, deriving extinction measures from NIR color-magnitude
diagrams (CMDs) using the
known properties of the red-clump giants. This method is capable of
rendering extinctions to distances as far as 8 kpc in the
Galactic plane.

In lieu of empirical extinction measures, a three-dimensional Galactic
dust distribution model can be adopted from which extinction is
derived. Previous studies at high and mid galactic latitudes 
have used ``slab'' models, where a vertical dust density profile is
adopted. \citet{Chen99} have extended the usefulness and accuracy of
such a model to relatively low latitudes by renormalizing their model
to the SFD extinction map. Large-scale models that describe the
dust distribution over the entire Galactic disk have in the past been
axisymmetric, correlating the radial variation of the dust
distribution with that of the gas (hydrogen) surface density, derived
from HI, HII and CO observations together with 
assumed values and gradients for the gas:dust (metallicity) and
CO:H$_2$ ratios \citep[eg.][]{Sky92, Ortiz93}. 
Recently models have been constructed based upon FIR data, where
Galactic emission is dominated by the thermally 
radiating dust. Initially these models were also axisymmetric
\citep{SMB97,davies97}, but more recently have adopted nonaxisymmetric
structures to account for their observed FIR emission, including the
Galactic warp and spiral arms \citep{DS01, LZM02}.

Here we present for general use a large-scale 
three-dimensional model of Galactic
extinction based on the Galactic dust distribution model of
\citet[][hereafter DS01]{DS01}. The following
two sections describe the construction and application of the
extinction model. In section four we describe how empirical extinction
measures can be derived from NIR CMDs, 
and we compare the predictions of the
model with these and other extinction measures. The summary section
details a number of caveats that a potential user of the model should
be aware of.

\section{Galactic extinction model}

Given a three-dimensional model of the dust distribution, $\rho_{\rm d}({\bf
x})$ the extinction in a given direction can be found by integrating along a
line-of-sight to the point of interest,
\begin{equation}
A_{\rm V}({\bf x}) = 1.086 \tau_{\rm V} = 1.086 \kappa_{\rm V} \int_0^{\bf x} \rho_{\rm d} ~ds,
\label{extx}
\end{equation}
while the total Galactic extinction is found by integrating to infinity.
The model of the dust distribution is detailed in DS01. We mention
here that this model is composed of three structural components: a
warped, but otherwise 
axisymmetric disk with a radial temperature gradient, spiral arms as
mapped by known HII regions, and a local Orion-Cygnus arm segment.
The structural parameters of the dust model are constrained by the FIR
observations of the COBE/DIRBE instrument, while
the determination of the parameter $\kappa_{\rm V}$, and the
decomposition of the flux density into dust density and
emissivity, is achieved by modeling the extinction in the COBE/DIRBE NIR
observations. However, the model as presented here has been amended in its
description of the spiral arm geometry.

\begin{figure}[t]
\epsfxsize=6in
\epsffile{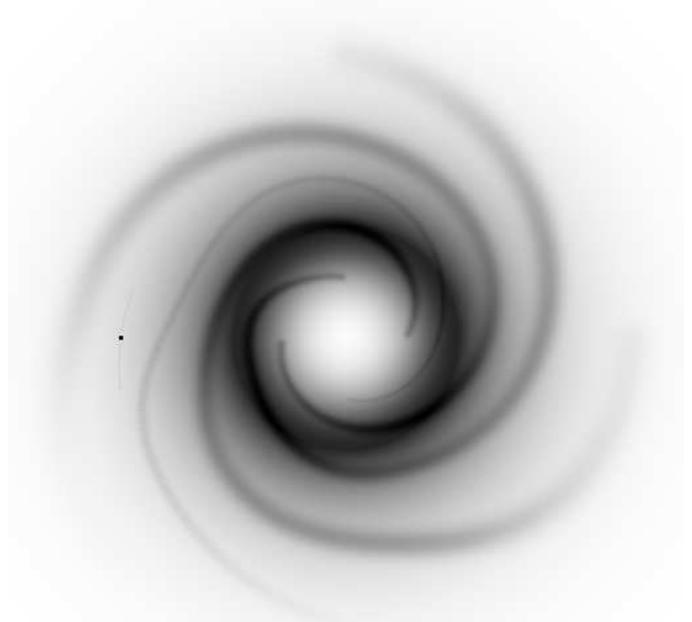}
\caption{Map of the dust surface density from the dust distribution model
with the extended geometry for the spiral arms. 
The Sun's location is indicated by the point at center-left.
}
\label{dmap}
\end{figure}

The previous spiral arm geometry used in DS01 was based on the mapping
of Galactic HII regions \citep{GG76, TC93}. Due to a 
lack of data, this map did not extend to the opposite side of the Galaxy.
In order to provide a realistic model of the spiral
arm contribution over the whole of the Galaxy, the
extended geometry of \citet{BM02} is adopted.
Fig. \ref{dmap} shows the dust surface density of the model,
including the new spiral arm component.
No other parameters of the dust model have been changed, and in fact the
extension of the arms does not lead to any significant differences in
the FIR emission profiles as predicted by the model of DS01, because
the spiral arms on the far-side of the Galaxy are unresolved in
the FIR DIRBE data. This model will need further refinement
once FIR emission toward the Galactic center is considered, as it
predicts arm tangents within $|l| < 20^\circ$ which are not evident
in the Galactic plane FIR emission profile.

Another feature of the dust model of DS01 is the use of
direction-dependent rescaling factors that are based on the FIR
residuals between the DIRBE 240 $\mu$m data and the predicted emission of
the parametric dust distribution model, and effectively ``correct''
the dust column density of the smooth model to account for small
(angular) scale structure not described explicitly by the model.
In practice, for any given DIRBE pixel, one of the three structural
components is rescaled to reproduce
the FIR flux. The component chosen for rescaling is typically that
which needs the least fractional change in its column density to
account for the FIR residual, though the spiral arms are
preferentially chosen near the Galactic plane (see DS01 for
details). The effect of the rescaling factors is to add (detract)
dust along the entire line-of-sight {\em for the rescaled component
only}, therefore their application represents an approximate
correction to the model in the sense that no bias with respect to
distance is effected, apart from that described by the rescaled component.
It should also be noted that the rescaling procedure implicitly
assumes that the deviation between the predicted and observed FIR
emission is {\em not} due to variations in dust temperature.

In DS01 the rescaling factors were used to refine the dust model 
when accounting for extinction in
their modeling of the J and K band emission observed by COBE, hence
only applied for galactic latitudes $|b| < 30^\circ$. As a practical
matter, the rescaling factors
for latitudes $|b| > 30^\circ$ are based on the SFD Galactic extinction
map, \nocite{Schleg98} as the low signal-to-noise of the 240
$\mu$m emission at high galactic latitudes does not allow their
reliable determination.  
Rescaling to the SFD extinction map was done as follows:
The rescaled dust density can be expressed as
\begin{equation}
\tilde{\rho}_{\rm d} = \sum f_i \rho_i
\end{equation}
where the sum is over the structural components of the dust density
model ($i=$ disk, spiral arms, local Orion arm), and the rescaling factors
$f_i$ are direction dependent, i.e. $f_i(l,b)$. Rescaling is applied
to only one component, so that at most one of the factors $f_i \neq 1$
for a given line-of-sight. For any given direction the total extinction is then
\begin{equation}
\tilde{A}_{\rm V}(\infty) = \sum f_i A_i(\infty) ,
\label{totext}
\end{equation}
where
\begin{equation}
A_i(\infty) = 1.086 \kappa_{\rm V} \int_0^\infty \rho_i ~ds,
\end{equation}
the integral being taken along the line-of-sight, $\rho_i$
corresponding to the dust density associated with component $i$ and
$\kappa_{\rm V}$ is the mean opacity.  Rescaling to the
SFD extinction map, $A_{\rm SFD}$, is effected by insisting that
$\tilde{A}(\infty) = A_{\rm SFD}$, which leads to
\begin{equation}
f_{\rm disk} = \frac{A_{\rm SFD} - \sum_{i \neq {\rm disk}} A_i(\infty)}{A_{\rm disk}(\infty)}.
\end{equation}
Only the disk component need be rescaled, as at high Galactic
latitudes it is only this component that
contributes significantly to the dust column density.

\begin{figure*}[ht]
\centering
\epsfxsize=17cm
\epsffile{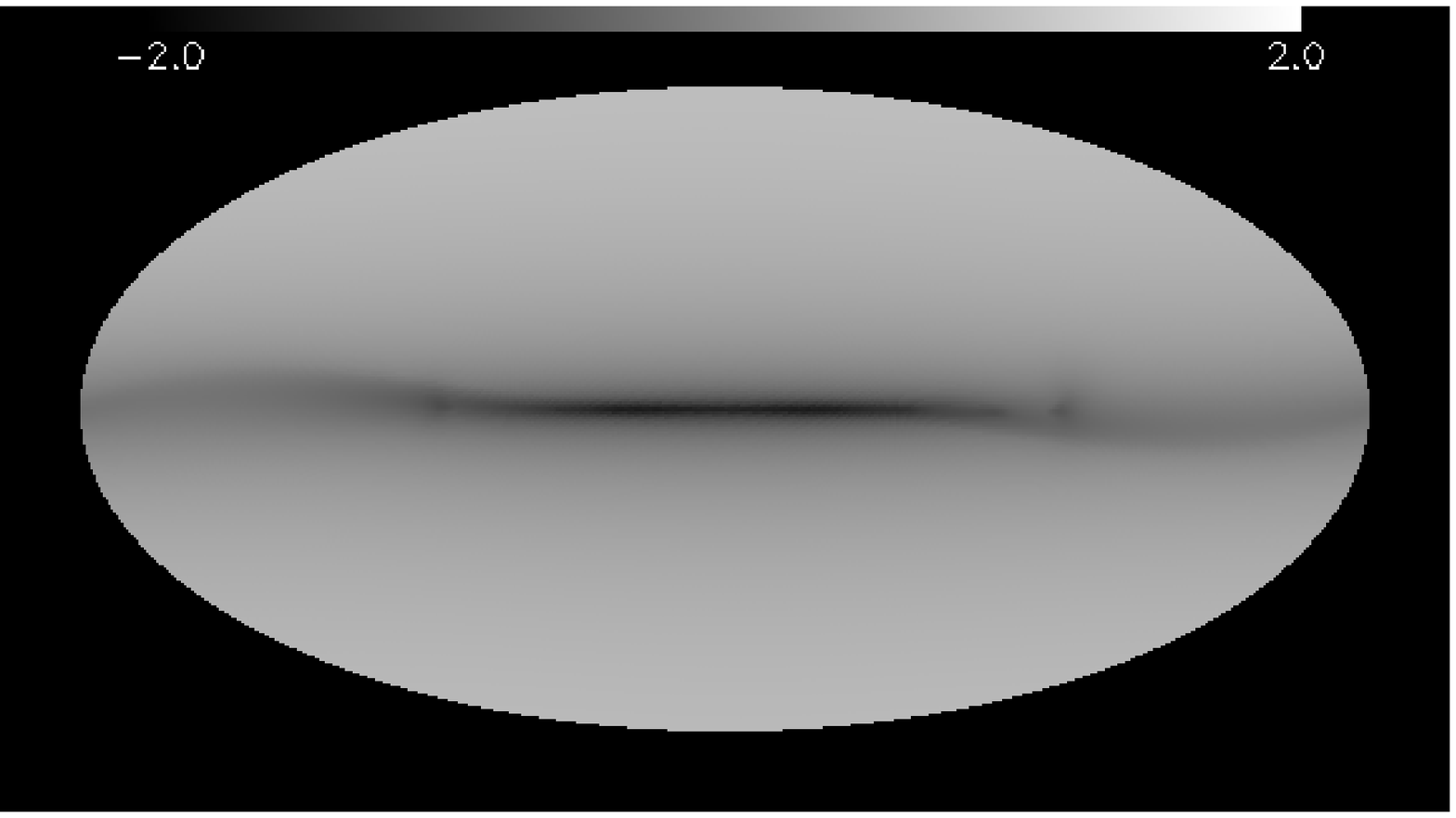}
\epsfxsize=17cm
\epsffile{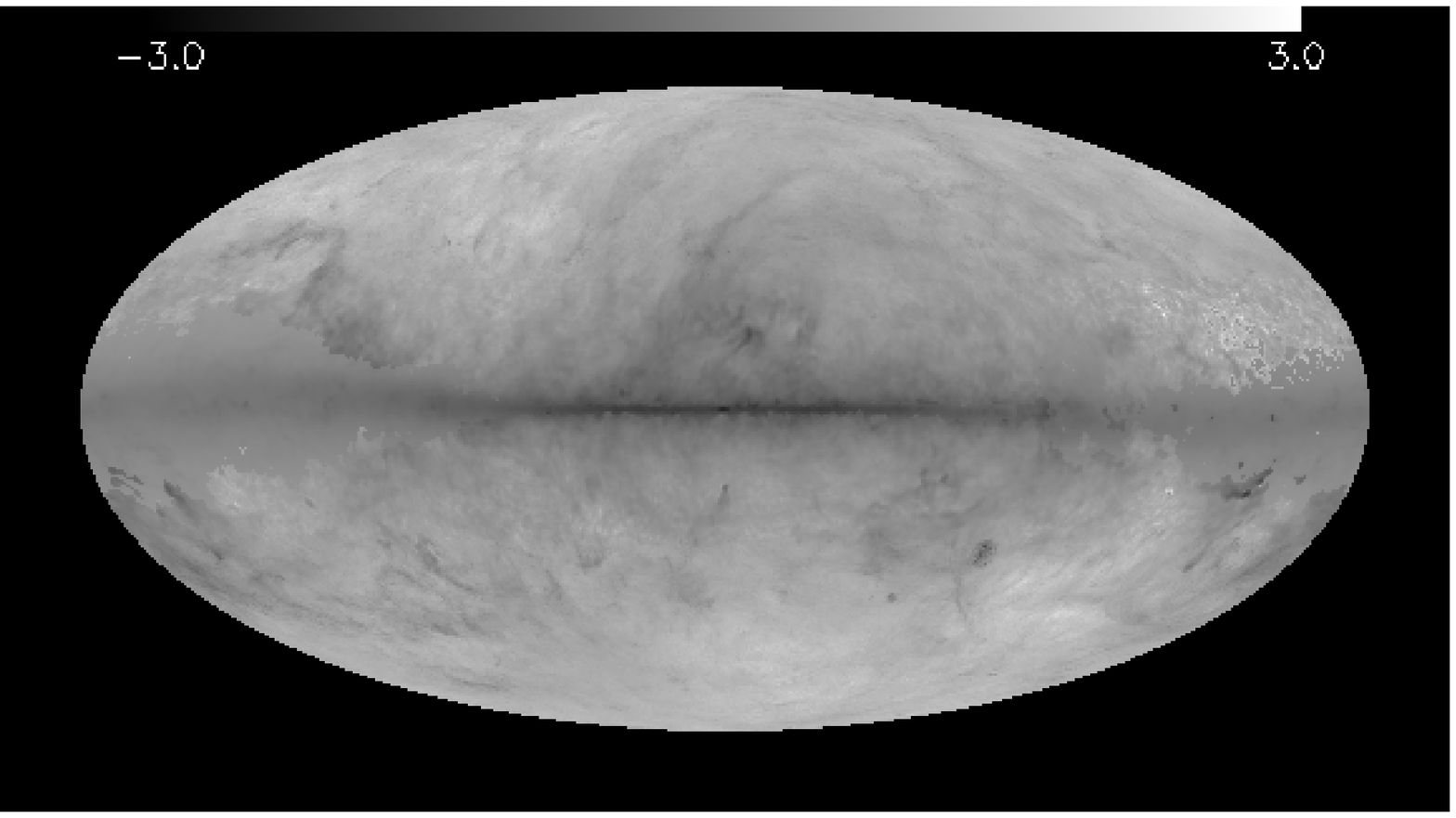}
\caption{
Sky maps of the predicted total Galactic extinction in a Mollweide
projection, without (top) and with (bottom) rescaling. The direction
to the Galactic center is at the center of 
the plot, positive galactic longitudes to the left. 
The quantity plotted in grey scale is $-\log (A_{\rm V})$.
}
\label{newext}
\end{figure*}

\begin{figure*}
\centering
\epsfxsize=17cm
\epsffile{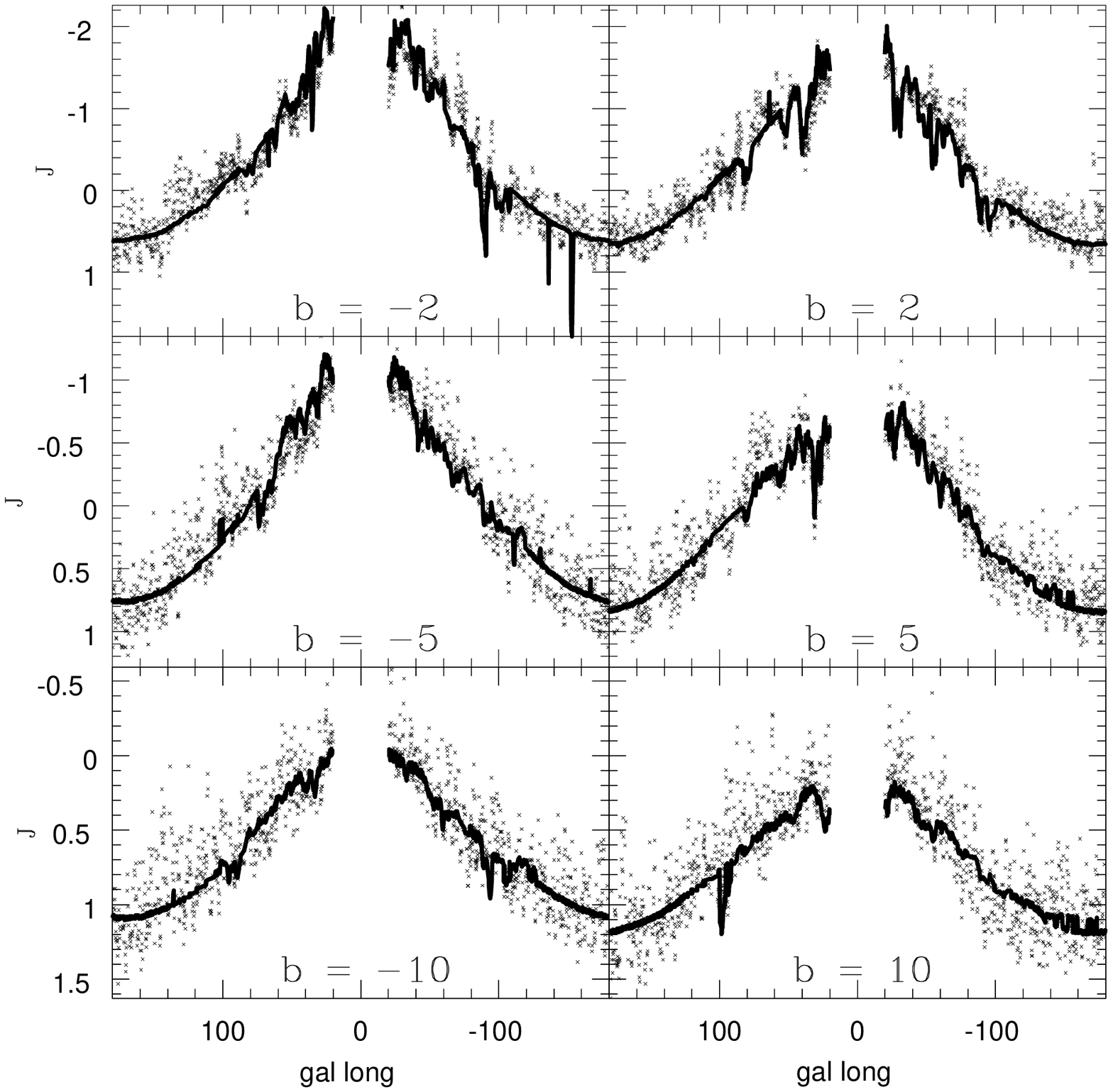}
\caption{J band emission profiles at low galactic latitudes, as
observed by the COBE satellite (crosses) and predicted by the dust
model using rescaling factors (from DS01).
}
\label{jprof}
\end{figure*}
\begin{figure*}
\centering
\epsfxsize=17cm
\epsffile{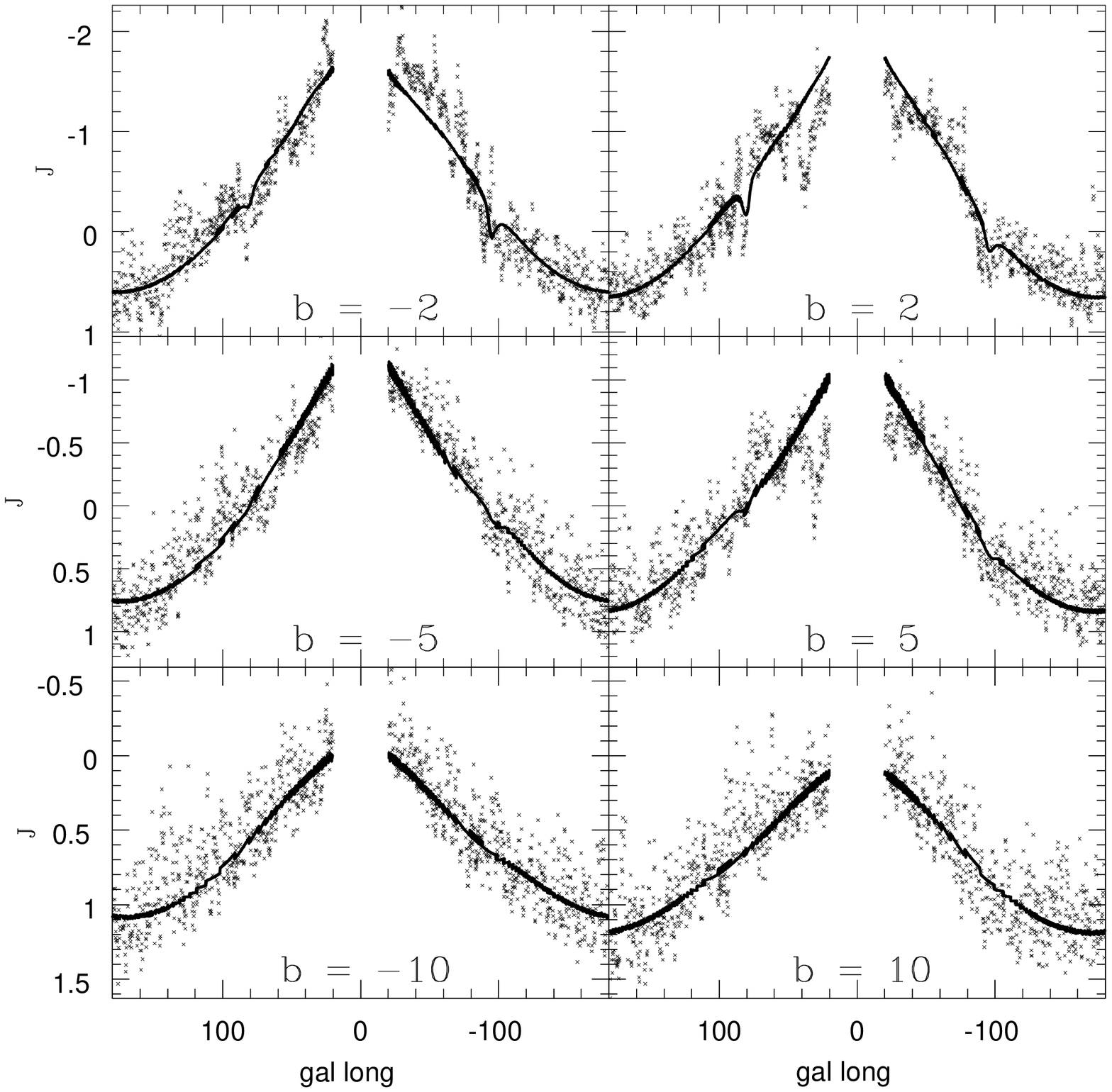}
\caption{As in previous figure, but using the dust model without
rescaling factors.
}
\label{jprof_nrsc}
\end{figure*}

Fig. \ref{newext} shows the skymaps of total Galactic
extinction, both with and without rescaling. 
Rescaling effectively adds angular detail that is not described by the
parametric dust model. 
To give the reader an appreciation of the importance of the rescaling
factors at low galactic latitudes, 
J band emission profiles are shown
in Figs. \ref{jprof} and \ref{jprof_nrsc} that both employ and
neglect their use. 

\section{Application}


An estimate of the extinction to any point within the Galaxy using the
rescaling factors can be achieved by either integrating the rescaled
dust model, $\tilde{\rho}_{\rm d}$, along the line-of-sight to the point of
interest (Eq. \ref{extx} with $\tilde{\rho}_{\rm d}$),or by summing
over the extinction 
contributions of each component, determined separately via a line-of-sight
integration, and applying the rescaling factor to the appropriate
component. However, such integrations are time
consuming, especially if many lines-of-sight need to be considered. To
provide a more efficient means of finding the extinction within the
Galactic disk we have constructed a set of three-dimensional cartesian
grids of extinction in $V$, each corresponding to a separate component of 
the dust model, as well as a table of the rescaling factors described
above. Both large-scale and small scale grids are provided; the former
cover the entire Galactic disk while the later describe the extinction
about the Sun in greater detail. 
Using the three-dimensional extinction grids a value for the
extinction $A_i$ due to each of the components $i$ of the dust
distribution can be found for any point in the Galactic disk via
interpolation. Together with the appropriate rescaling
factor a final estimate of the extinction is arrived at:
\begin{equation} 
\tilde{A}_{\rm V}(l,b,r) = \sum f_i(l,b) A_i(l,b,r).
\label{tota}
\end{equation}
Trilinear interpolation from the grids described above is orders of
magnitude faster than numerically integrating the dust model to 
individual sources, and the interpolation error is of the order
of 0.2\%.
 
\begin{figure}[t]
\epsfxsize=3.5in
\epsffile{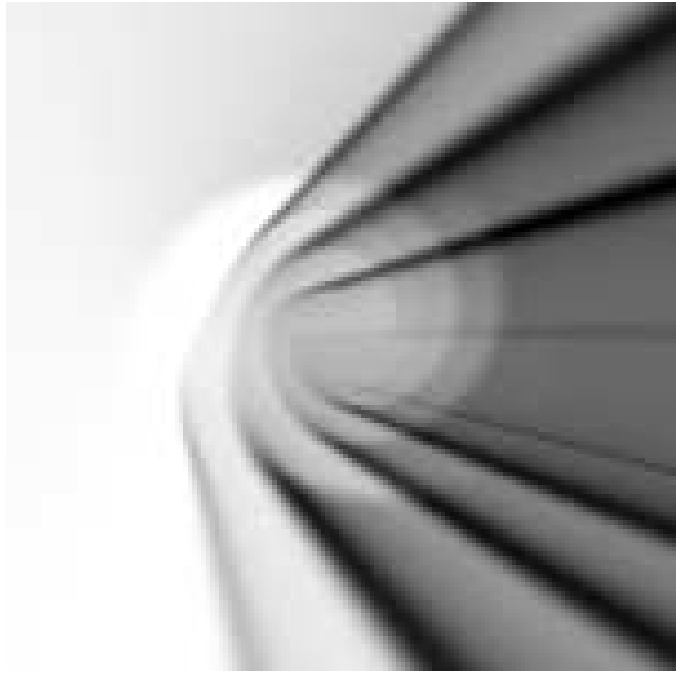}
\caption{
Extinction in the Galactic plane due to the spiral component of the
dust distribution out to $R = 15$kpc from the Galactic center. 
The Galactic center is at center, with the Sun
located at center-left. The tangents to the spiral arms are clearly
evident. The maximum extinction is 5.7 magnitudes in V.}
\label{avspir}
\end{figure}

As an example, Fig. \ref{avspir} shows a slice in the Galactic plane 
through the large-scale grid for the spiral arm component. Both this
grid and that of the disk component have $xyz$ dimensions of $30 \times 30
\times 1$kpc, the $z=0$ plane corresponding to the Galactic plane.
Thus they describe the extinction out to a galactocentric
radius of $R = 15$kpc. The grid spacing parallel to the Galactic plane
is 200 pc, while perpendicular to plane the resolution is 20pc.
Meanwhile, the large-scale grid for the Orion arm component has reduced
dimensions and finer grid spacing parallel to the Galactic plane, so
that it covers only part of the Galaxy, having dimensions of $3.75
\times 7.5 \times 1$kpc.  

%

To provide greater detail near the Sun and to avoid interpolation
errors, two local grids are constructed, one for the Orion arm and another for
the disk component; the spiral arm component does not make a
significant contribution near the Sun. Both grids have their $xy$
coordinates centered on the Sun, 
though with different grid spacing in the $xy$ directions.
(The grid spacing in the $z$ direction for all grids is
$\Delta z = 0.02$ kpc.)
The second of these local grids, that of the disk component, is
necessary only to avoid interpolation errors which would otherwise be greater
than 5\% for heliocentric distances less than 500 pc with the larger grid
spacing of the large-scale grid. Ideally this region about the Sun
should be described by a more detailed local model of the dust
distribution.

The rescaling factors over the entire sky are also supplied on the
same resolution as the FIR COBE/DIRBE data:
Because the rescaling factors are based on the residuals between the DIRBE FIR
observations and the predictions of the dust distribution model, the
rescaling factors are rooted to the DIRBE data 
structure; for each DIRBE pixel there is an associated rescaling
factor.  This introduces some complication in the retrieval of the
rescaling factors for any arbitrary direction as the DIRBE sky maps 
use a nonstandard projection and a binary pixel ordering scheme.
(See http://space.gsfc.nasa.gov/astro/cobe/skymap\_info.html for a
brief introduction and Appendix G of the COBE Guest Investigator
Software (CGIS) Software User's Guide (version 2.2), retrievable at 
ftp://rosette.gsfc.nasa.gov/pub/cobe-gi/doc/, for further details.)
To apply the rescaling factors the user must determine the DIRBE pixel that
corresponds to the direction of interest. There are at
least three options for this: 
\begin{enumerate}
\item Write a routine that searches through the entire list of pixel
coordinates for the nearest pixel.
\item Work within the {\small IDL} environment developed by the COBE
data reduction team which provides an efficient
means, via CGIS {\small IDL} code, to retrieve the pixel number
nearest a given direction. 
This {\small IDL} package with instructions for installation can be
found at 
http://space.gsfc.nasa.gov/astro/cobe/cgis.html.
\item Use stand alone CGIS standard {\small FORTRAN} code to
retrieve the pixel number for a given direction. This code can be
downloaded via anonymous ftp from ftp://rosette.gsfc.nasa.gov/pub/cobe-gi/. 
\end{enumerate}
Thus a file is provided for the user containing, for each DIRBE pixel,
the DIRBE pixel number, its galactic coordinates, an index of the
component to be rescaled and its rescaling factor. This file, together
with those containing the extinction grids and detailed
instructions for their use, can be found at the anonymous ftp site
ftp://ftp.to.astro.it/astrometria/extinction/.

To summarize, an algorithm for finding the extinction $A_{\rm V}$ to a
point in the Galaxy $(l,b,r)$ is outlined below:
\begin{itemize}
\item Find the COBE pixel coinciding to $(l,b)$.
\item Recover the rescaling factor $f_i$  and component $i$ to be rescaled.
(Set $f = 1$ for the other components.)
\item Determine the grid coordinates:\\
$(l,b,r) \longrightarrow (x,y,z) \longrightarrow (i,j,k)$
\item Interpolate from appropriate grids to find $A_i(l,b,r)$ for each
component.
\item Sum over the components (Eq. \ref{tota}) to arrive at
$\tilde{A}_{\rm V}(l,b,r)$. 
\end{itemize}
An example of {\small IDL} and {\small FORTRAN} code that performs the above
algorithm can be found on the anonymous ftp site mentioned above.

\begin{figure}[t]
\epsfxsize=3.5in
\epsffile{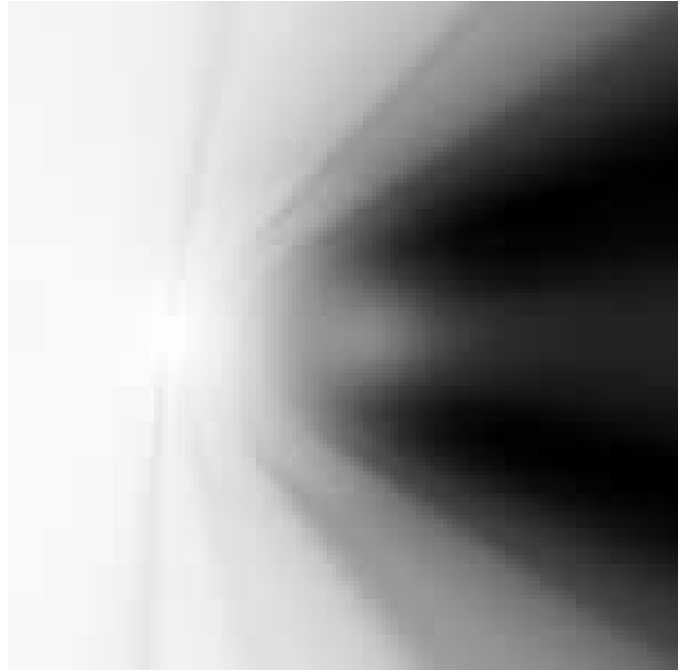}
\caption{
Extinction map in the Galactic plane resulting from all components of
the dust model (disk, spiral arms and the local Orion-Cygnus arm).
Orientation and scale is as in the previous diagram and the maximum
extinction is 37 magnitudes in V.
}
\label{avgrid}
\end{figure}

It is important to note that if one decides {\em not} to use the
rescaling factors it is sufficient to use only two grids,
a large-scale and a local grid, which are each a sum of the grids
mentioned above, and perform only a single interpolation from the
appropriate grid. Fig. \ref{avgrid} shows the resulting large scale
grid, also provided to the user.
One might wonder why the same is not done above, that is, construct a single
three-dimensional grid of the final rescaled $\tilde{A}_{\rm V}(i,j,k)$. 
The reason is that the
use of the rescaling factors would introduce discontinuities in such a
grid, rendering interpolation unreliable; first interpolation must be
done on each (smooth) component, then the direction dependent
rescaling factor applied. 

{ \bf As a final detail we mention that the extinction from the
model, given in the V band, can be transformed to other wavebands using
the $(A/A_{\rm V})$ ratios given by 
\citet{RL85}. Recently there has been some debate whether these ratios
may be too high in the NIR \citep{Glass99,Draine03}, however these ratios are 
preferred here in order to remain consistent with
the NIR modeling of DS01. It should also be noted that if using these
ratios one assumes a specific reddening curve which in practice is
spatially variable blueward of the Johnson R band
\citep{Mathis90,Fitz99}.  
}

\section{Comparing model with data}

The Galactic extinction map of SFD \nocite{Schleg98}, $A_{\rm SFD}$, 
and of the extinction model, $A_\infty$, are both based in
part on COBE FIR data, however the maps were derived using different
approaches. It is therefore interesting to compare the two extinction maps
{\em before} using the SFD map to rescale the high latitude data. The
two maps of Galactic extinction were compared for each COBE pixel
between galactic latitudes $5^\circ < |b| < 40^\circ$. The lower
boundary is the limit of reliability for $A_{\rm SFD}$ as stated by
\citet{Schleg98}, while above $|b| = 40^\circ$ the 240$\mu$m data
becomes too noisy to derive reliable rescaling factors. In addition
data in the directions of the Orion nebula, Rho Ophiucus, Andromeda
and the LMC and SMC were removed. The mean residual of the remaining data,
$\langle A_{\rm SFD} - A_\infty \rangle$, was found to be 0.096
magnitudes, while the mean relative difference, $\langle (A_{\rm SFD}
- A_\infty)/A_{\rm SFD} \rangle$, 
was equal to 0.024, showing that the two maps have nearly the same
normalization. This concordance between the two extinction maps is the
reason why the boundary for the different rescaling schemes at $|b| =
30^\circ$ is hardly evident in Fig. \ref{newext}. However, the mean {\em
absolute} residual, $\langle |A_{\rm SFD} - A_\infty| \rangle$ has a
value of 0.15 magnitudes, or a mean relative absolute difference of 
0.19, showing that there is substantial scatter in the difference
between the two maps.  


However, rather than provide a sky map of the total Galactic
extinction, the primary purpose of the extinction model is to give
an estimate of extinction to any point {\em within} the Galactic
disk. To evaluate the model's performance in this regime we compare
extinction estimates from the model with empirical extinction measures
using NIR data from the second incremental data release (2IDR) of the
2MASS\footnote{2MASS is a joint project of the Univ. of Massachusetts
and the Infrared Processing and Analysis Center, funded by NASA
and NSF.}  project (\citet{2MASSrel2,2MASSpreCat},  
http://www.ipac.caltech.edu/2mass/releases/docs.html). 

In \citet[][hereafter L02]{L02} a method was presented to obtain both
the star density and interstellar extinction along a line-of-sight by
extracting the well-known red-clump population (spectral type K2III) 
from the infrared color-magnitude diagram (CMD). These stars 
constitute the majority of the disc giants (Cohen et al. 2000;
Hammersley et al. 2000) and can be easily identified in the NIR CMDs.
The method
is extensively described in L02, so we give only a brief summary with
some additional details here. 
 
\begin{figure}[t] 
{\par\centering 
\resizebox*{6.7cm}{6.7cm}{\epsfig{file=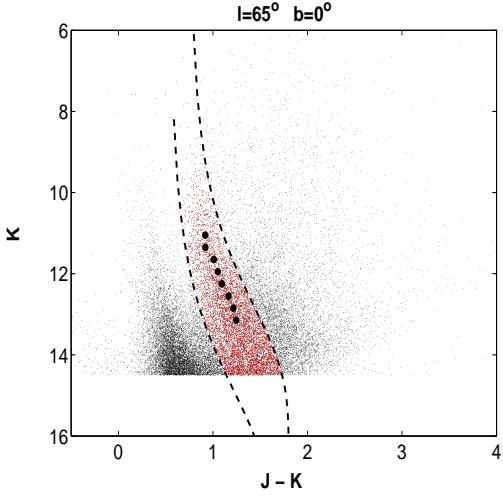,height=6.7cm}}\par} 
\caption{NIR color-magnitude diagram for a typical low-latitude field
($l=65^\circ$, $b=0^\circ$), taken from the 2MASS 
survey. Dashed curves delineate the selected region isolating the
red-clump giants. Filled  
circles show the maxima of the red-clump for individual magnitude bins.} 
\label{Fig:CM1} 
\end{figure}

(${\rm J-K},m_{\rm K}$) CMDs are built for 1$^\circ$ x 1$^\circ$ fields by
using available 2MASS data. In these diagrams stars of the same 
spectral type (i.e. the same absolute magnitude) 
will be placed at different locations in the CMD; 
the effect of distance and extinction
cause the red-clump giants to form a broad diagonal branch running from
top left to bottom right in the CMDs. 
In order to isolate the red-clump sources in the CMD we use theoretical traces
of different spectral types, based on the updated ''SKY'' model \citep{Sky92},
to define the possible color range of  
the K-giant branch in the CMDs, without any further implication in the 
method. (The ``SKY'' model uses a double exponential disk for the dust 
distribution, used here to roughly approximate interstellar extinction.)   
For each field appropriate traces are chosen to 
isolate the K-giants and avoid contamination
by other stellar populations, especially dwarf stars
and M-giants (see Fig. \ref{Fig:CM1}).  
 
\begin{figure}[t] 
{\par\centering 
\resizebox*{6.7cm}{6.7cm}{\epsfig{file=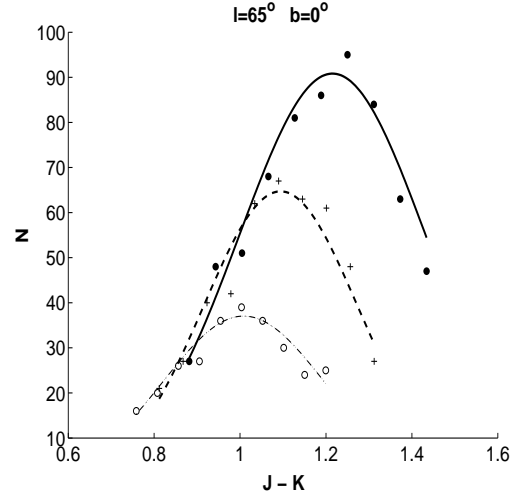,height=6.7cm}}\par} 
\caption{Gaussian fits (lines) to the red-clump counts (points) in
three magnitude bins for  
the field $l=65^\circ$ 
$b=0^\circ$. 
Solid line (filled circles) corresponds to the 12.7$<$m$_{\rm K}$$<$13
bin, the dashed line  
(pluses) to the 12.1$<$m$_{\rm K}$$<$12.4 bin, and the dot--dashed line
(open circles) to the   
11.5$<$m$_{\rm K}$$<$11.8 bin} 
\label{Fig:CM2} 
\end{figure}

Once the optimal traces have been selected, the giant stars are
extracted from the CMD and binned in apparent K magnitude, $m_{\rm
K}$. For each magnitude bin, count histograms in color are constructed. 
A Gaussian function was then fit to the histograms 
to determine the color of the peak counts at each magnitude:
\begin{equation}
f(x;A,\mu,\sigma) = A \exp[-(x-\mu)^2/2 \sigma^2],
\end{equation}
where $x$ is the binned (J$-$K) color and $\mu$ is taken as the color
of the peak. 
Fig. \ref{Fig:CM2} shows the Gaussian fits to the color histograms  
obtained at three different magnitude bins for the field of
Fig. \ref{Fig:CM1}. The maxima are identified as corresponding to the
mean observed color (J$-$K)$_{m_{\rm K}}$ of the K2III  
stars at the given apparent magnitude $m_{\rm K}$ 
since they are by far the most prominent population in the
selected region. (See also Fig. 2 in L02 for further details). 
 
The extinction A$_{\rm K}(m_{\rm K})$ can be determined by tracing how
the peak (J$-$K) of the red-clump counts changes with $m_{\rm K}$ and the
intrinsic mean color (J$-$K)$_0$ of the red-clump. From the color excess
and after \citet{RL85}:
\begin{equation} 
A_{\rm K}(m_{\rm K})=\frac{\rm (J-K)_{m_{\rm K}}-(J-K)_{0}}{1.52} .
\label{ak} 
\end{equation} 
Once the extinction $A_{\rm K}$ is estimated at the apparent magnitude
$m_{\rm K}$, a mean distance to the stars can then be found given
their mean absolute magnitude. For the K2III population the mean
absolute magnitude and the intrinsic color were  
assumed to be $M_{\rm K}=-1.65$ and (J$-$K)$_{0}$=0.75, with a Gaussian
dispersion of 0.3 mag in absolute magnitude and 0.2 in color (see
L02). Recent results in open clusters yield very similar values:
\citet{Groch02} obtained $M_{\rm K}=-1.62$ with a standard deviation
of 0.21 and mean color (J$-$K)$_{0}$ of around 0.7 for 
the old disc population, with small dispersion due to metallicity or
age gradients, while \citet{PGU03} show that  
the K-band magnitude of the red-clump is a good distance indicator,
independent of the metallicity and age, with some small dispersion in
the color dependence.  
  
Uncertainties of the extinction were estimated from the 
uncertainty of the mean color of the Gaussian component: 
\begin{equation} 
\sigma_{A_{\rm K}} = \sigma / (1.52 \sqrt{N}) 
\end{equation} 
where $\sigma$ is the standard deviation of the fitted Gaussian
and $N=A \sigma \sqrt{2\pi}$ is the estimated number of red-clump
stars. This uncertainty gives an indication  
of the statistical error only, and does not include other possible
systematic errors.  Some uncertainties of the method, such as
metallicity effects for the clump giants or differences between the
real red-clump color distribution and the assumed Gaussian function are fully
described in L02, so they will not be repeated here.  

Critical aspects that are to be considered in the Gaussian fit are
the possible contamination by dwarf stars and the effect of the
completeness limit of the  
2MASS survey. For fields such as those used here, dwarf contamination is
negligible for  
m$_{\rm K} \approx 13$ (see \S3.3 in L02 for details) while the 2MASS
survey is complete up to magnitudes  
of m$_{\rm K} \approx 14$. For this reason, the Gaussian fits have been
obtained in each field only up to m$_{\rm K} \approx 13$, 
well above the completeness limit of the survey and where the dwarf
contamination is small.  
 
This method has been used for near-plane regions in both the outer
and inner Galactic disc  
(L02; Lopez-Corredoira et al. 2003, in preparation) with very
satisfactory results. Some estimations  
of interstellar extinction have also been obtained in this way for
the star count
predictions of the \emph{Besan\c{c}on model} \citep{RC86A}
in the Galactic Plane and the results shown there are in good agreement
between the observed extinction in the K-band and the model
predictions \citep{Picaud03}. The main strength of  
the method is that it is empirical and we can extract the red-clump giants 
from the CMDs assuming only their mean absolute magnitude
and intrinsic color.  
 
%

\begin{figure*}[t]
\epsfxsize=3.5in
\epsffile{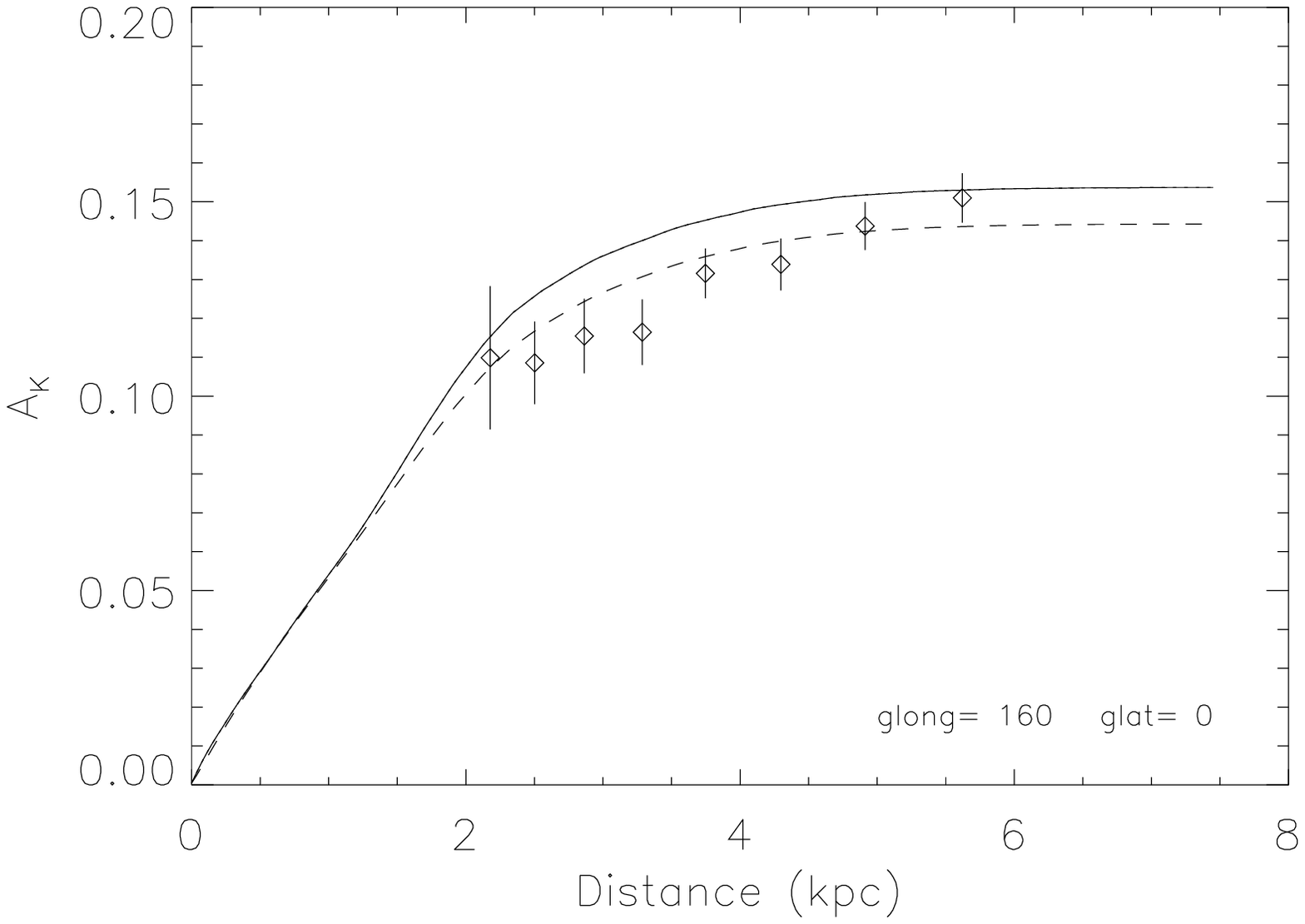}
\epsfxsize=3.5in
\epsffile{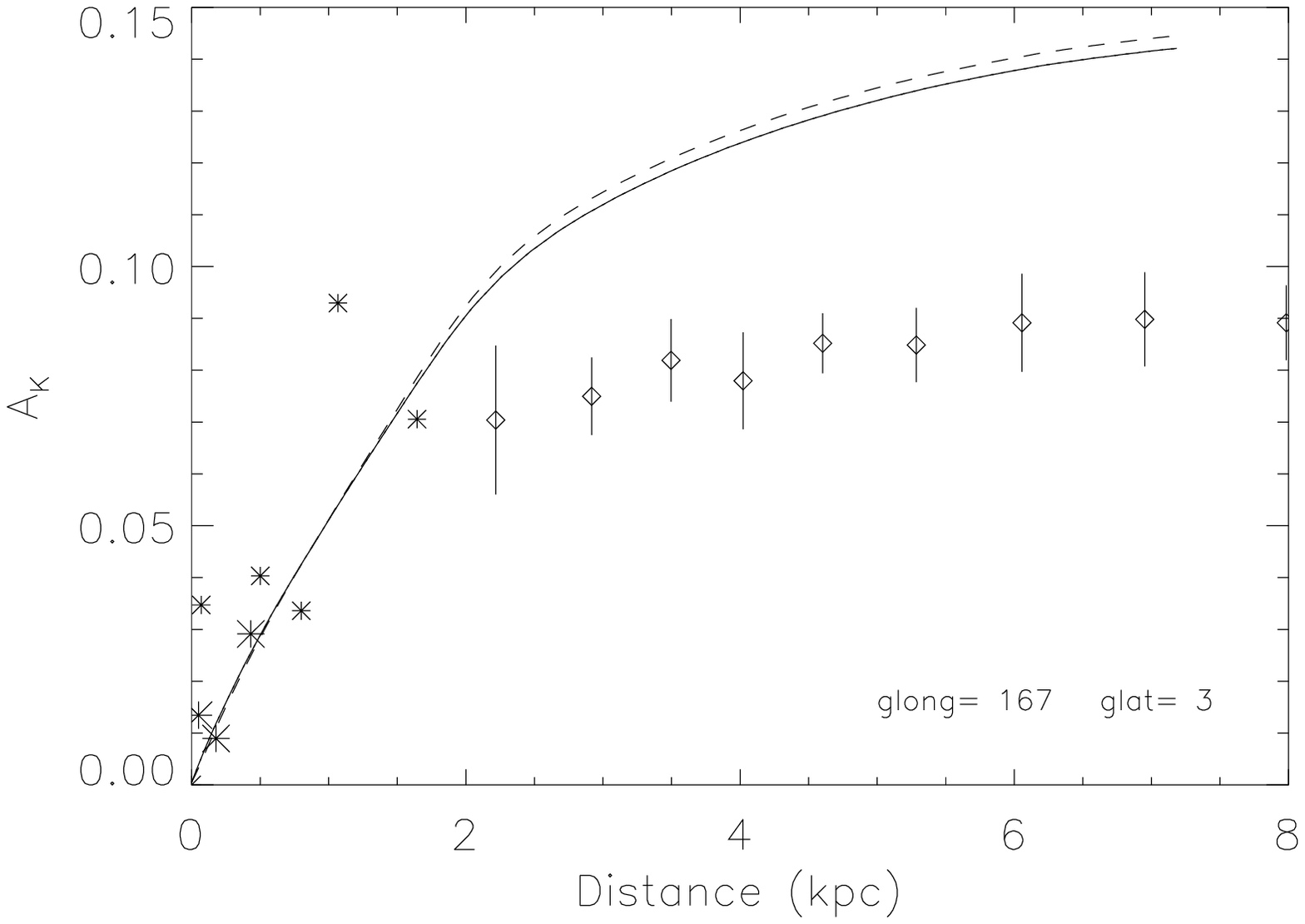}
\epsfxsize=3.5in
\epsffile{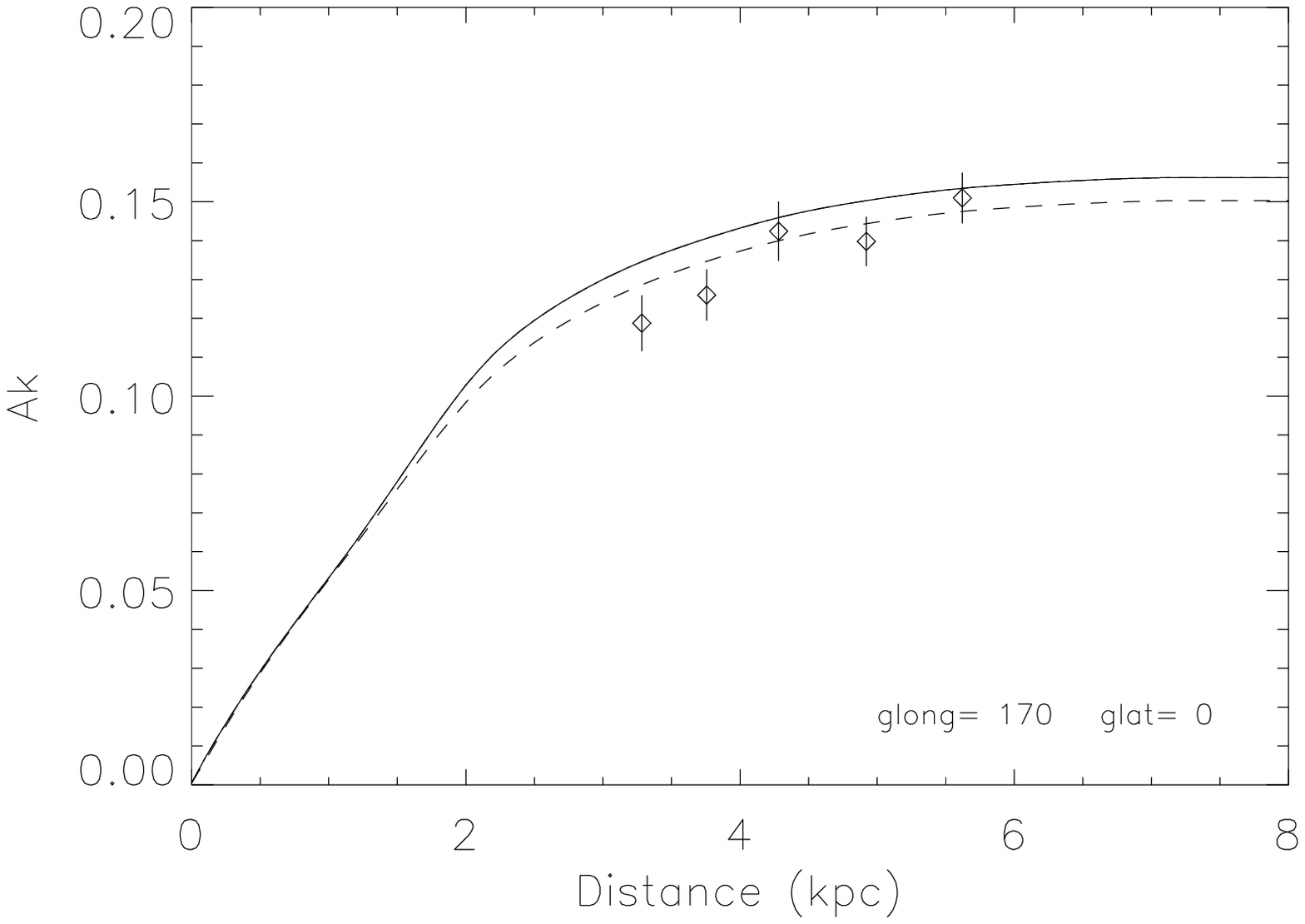}
\epsfxsize=3.5in
\epsffile{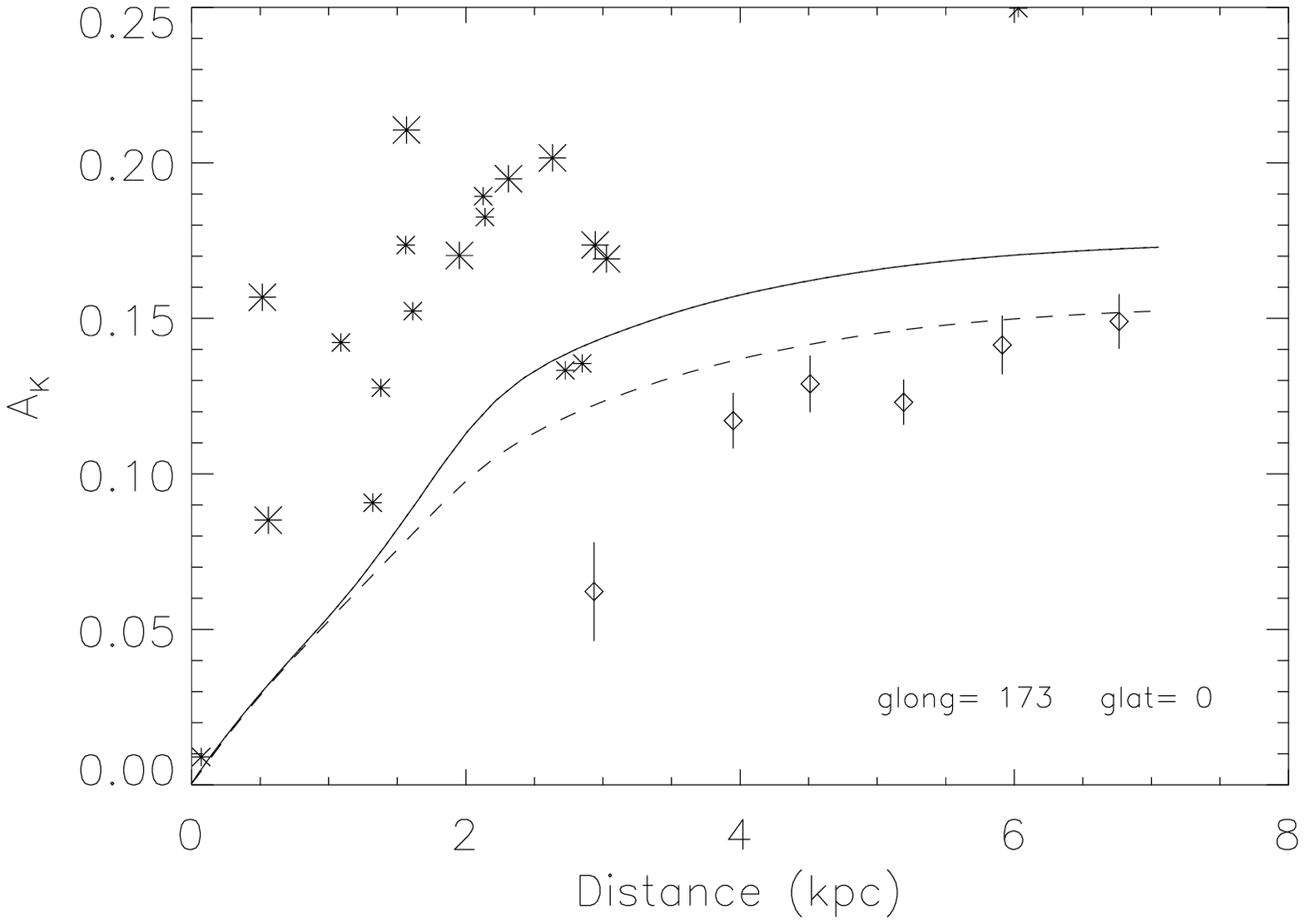}
\epsfxsize=3.5in
\epsffile{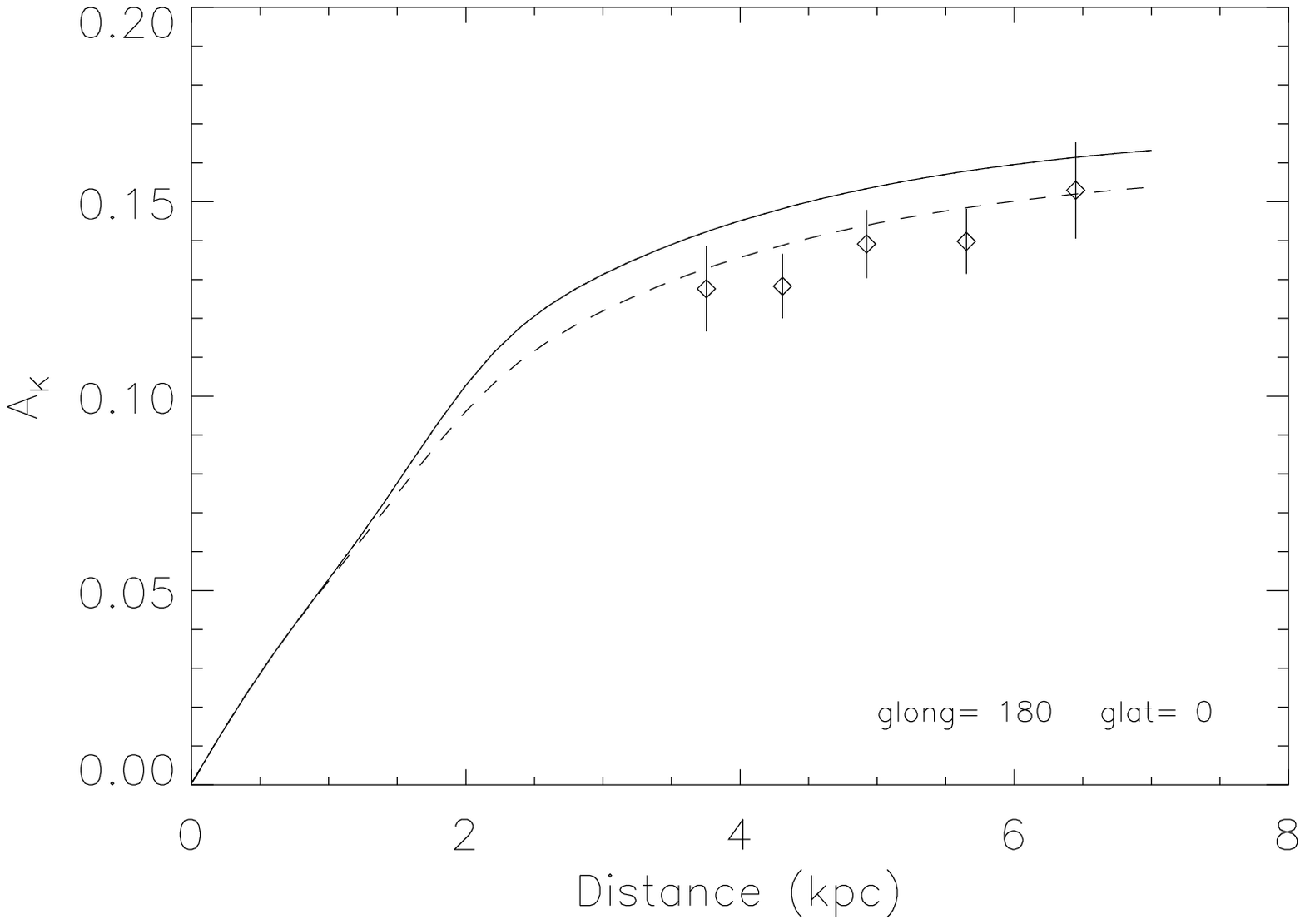}
\epsfxsize=3.5in
\epsffile{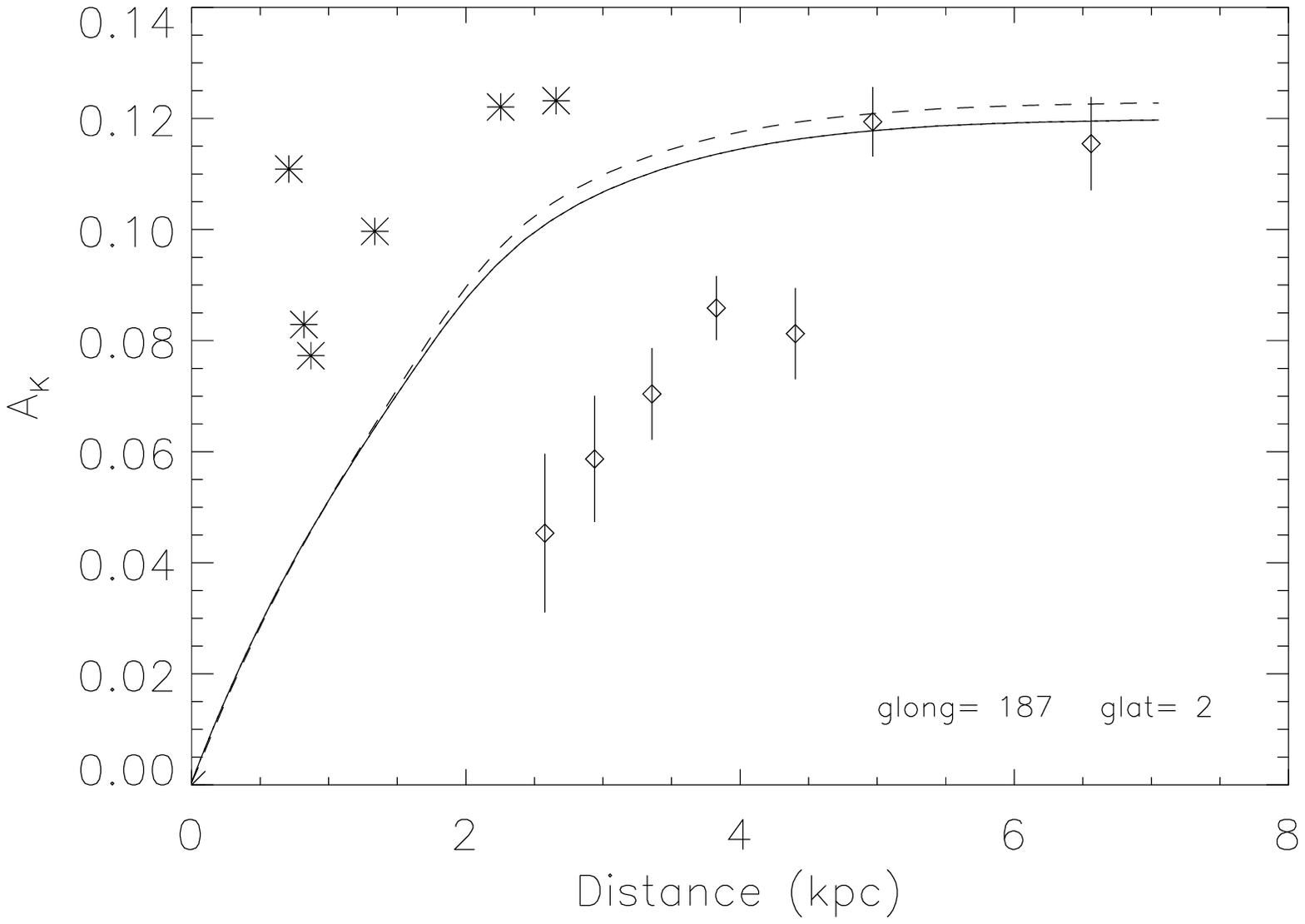}
\caption{
K band extinctions along lines-of-sight in the outer disk of the
Galaxy. Solid and dashed lines show the extinction according to the
model presented here, with and without rescaling factors
respectively. Diamonds show extinctions based on NIR CMDs of $1^\circ
\times 1^\circ$ fields using 2MASS data. Asterisks show extinction
measures from individual OB 
stars \citep{NKCat}, the large asterisks corresponding
to stars within $0.5^\circ$ of the directions indicated, small
asterisks stars within $1^\circ$. 
}
\label{outer}
\end{figure*}

\begin{figure*}
\epsfxsize=3.5in
\epsffile{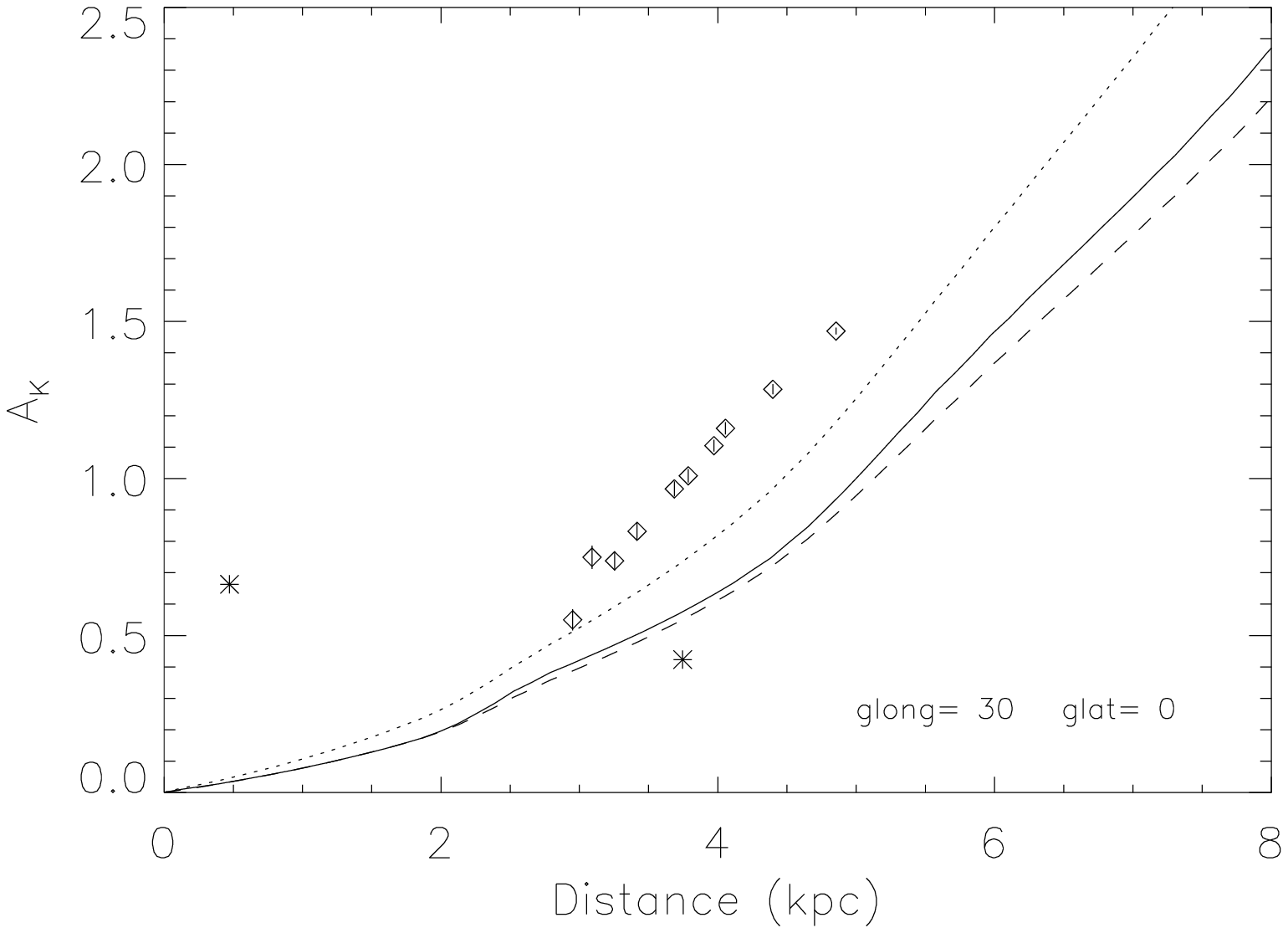}
\epsfxsize=3.5in
\epsffile{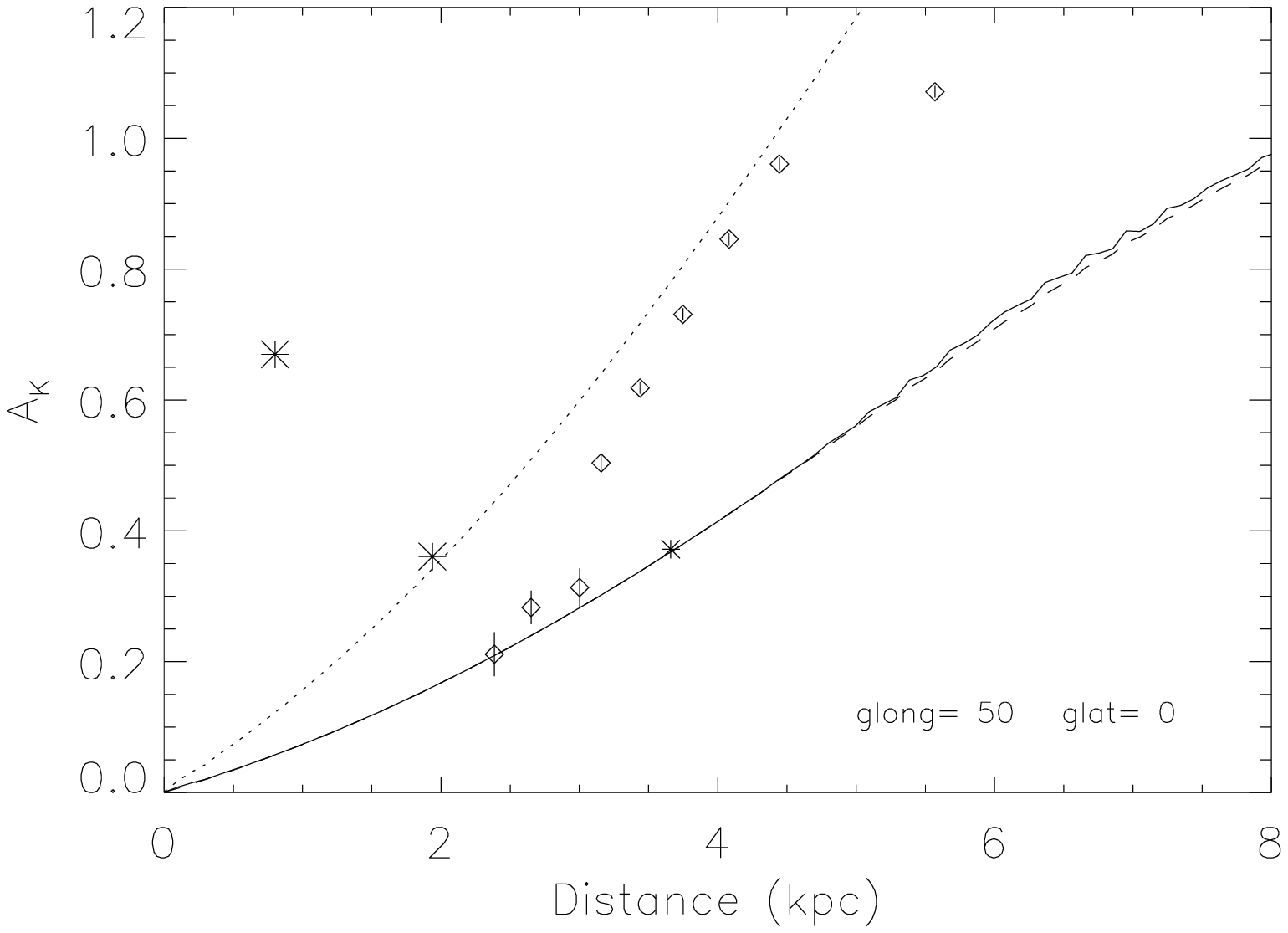}
\epsfxsize=3.5in
\epsffile{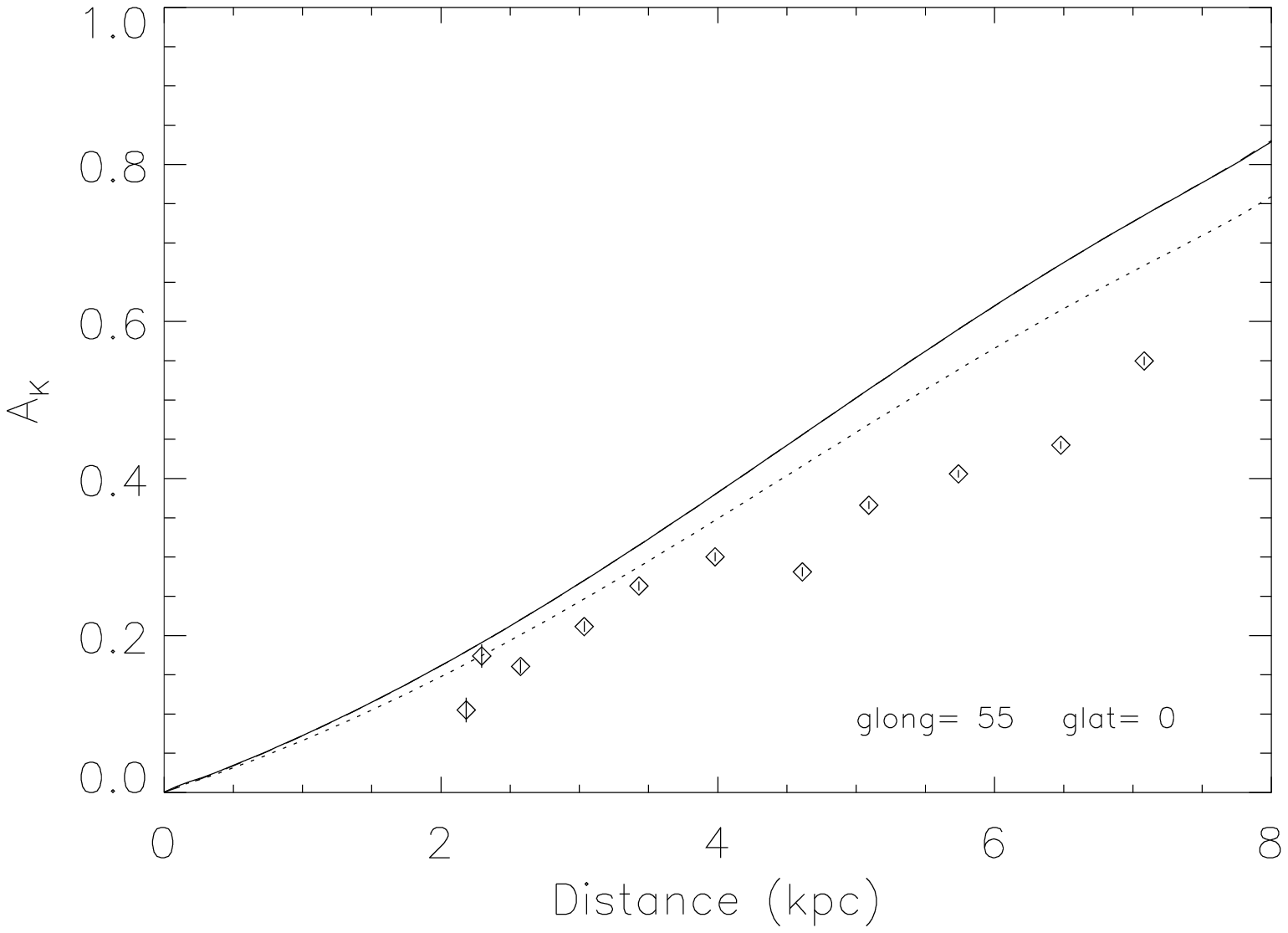}
\epsfxsize=3.5in
\epsffile{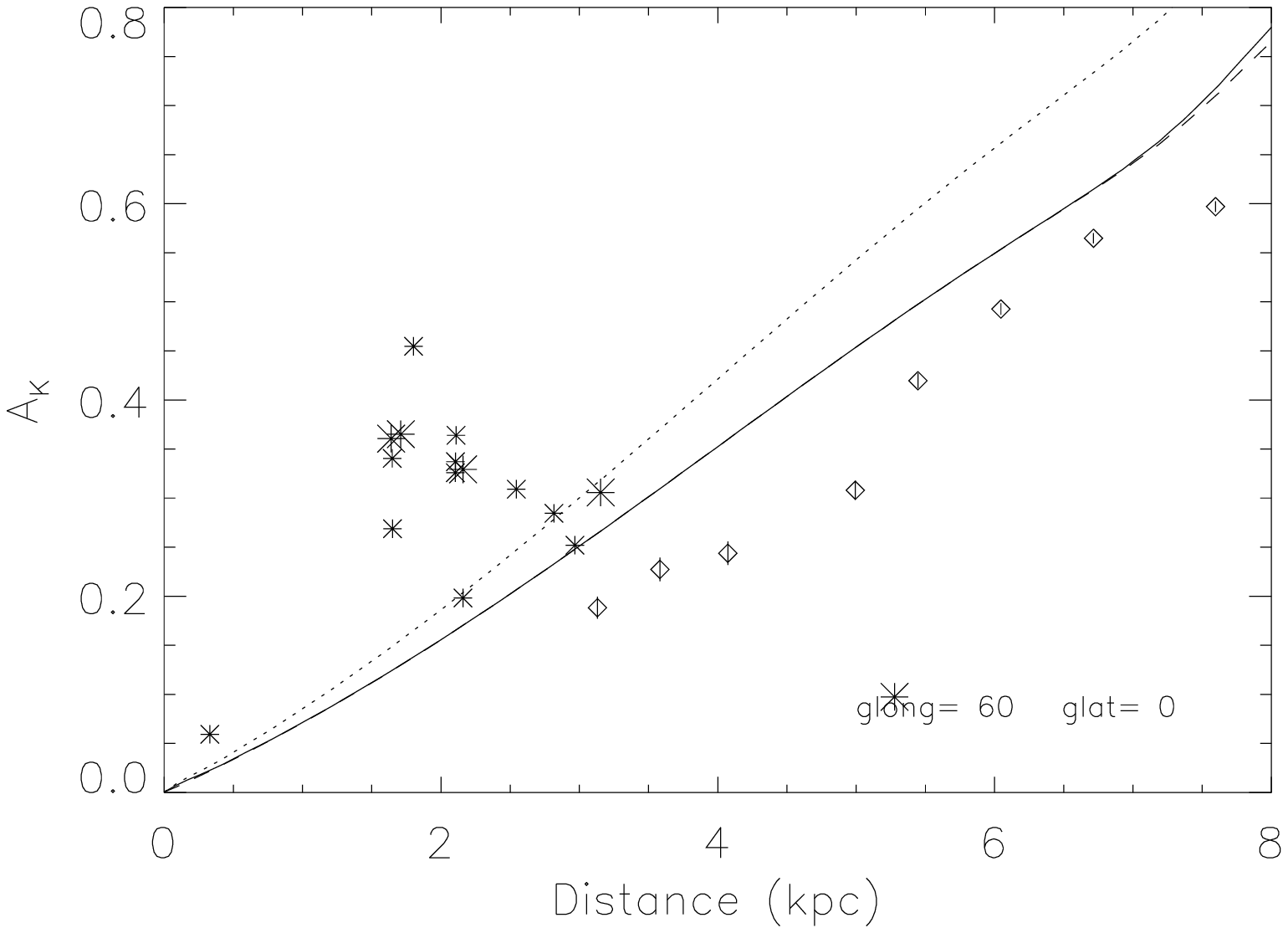}
\epsfxsize=3.5in
\epsffile{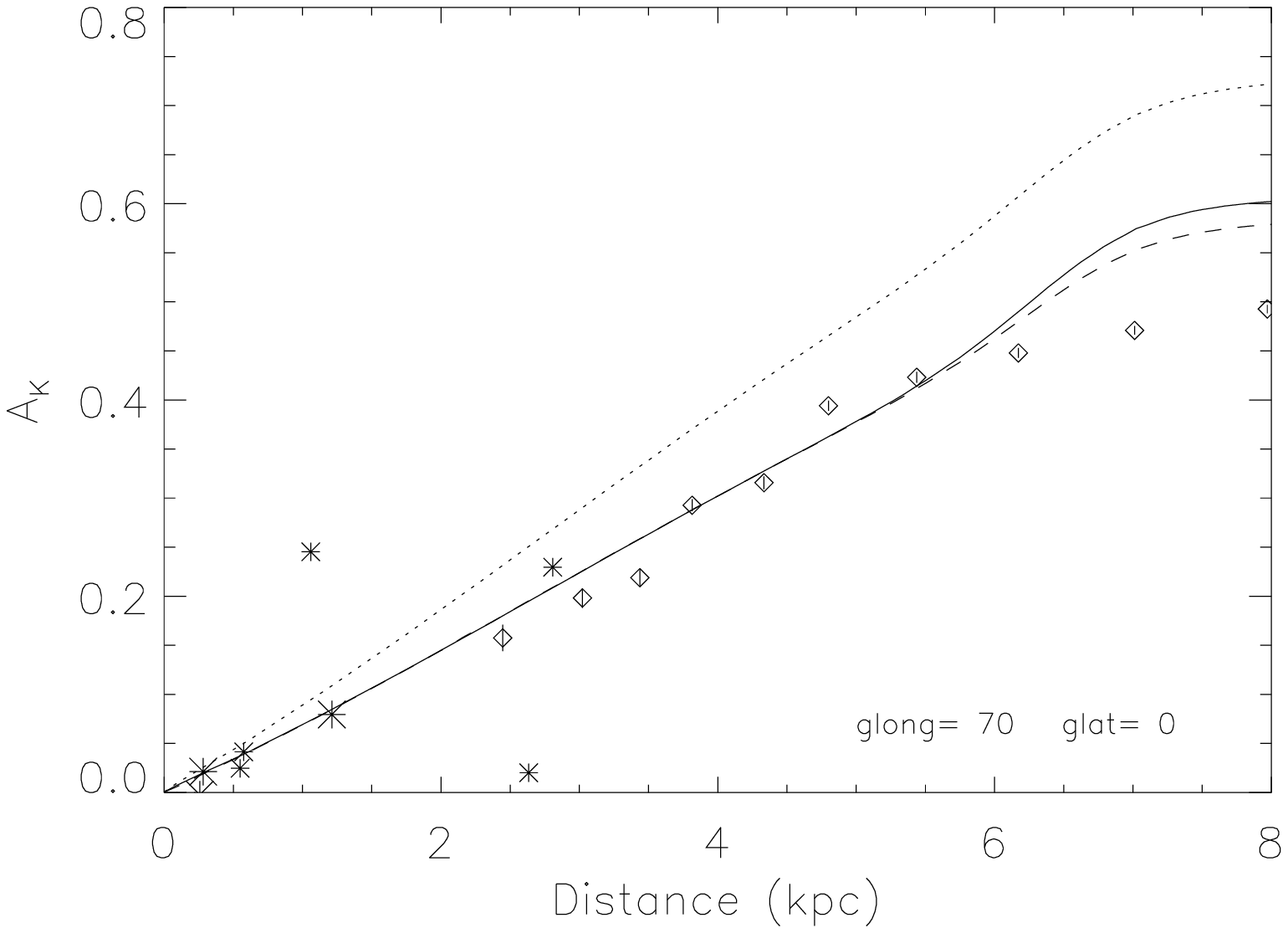}
\epsfxsize=3.5in
\epsffile{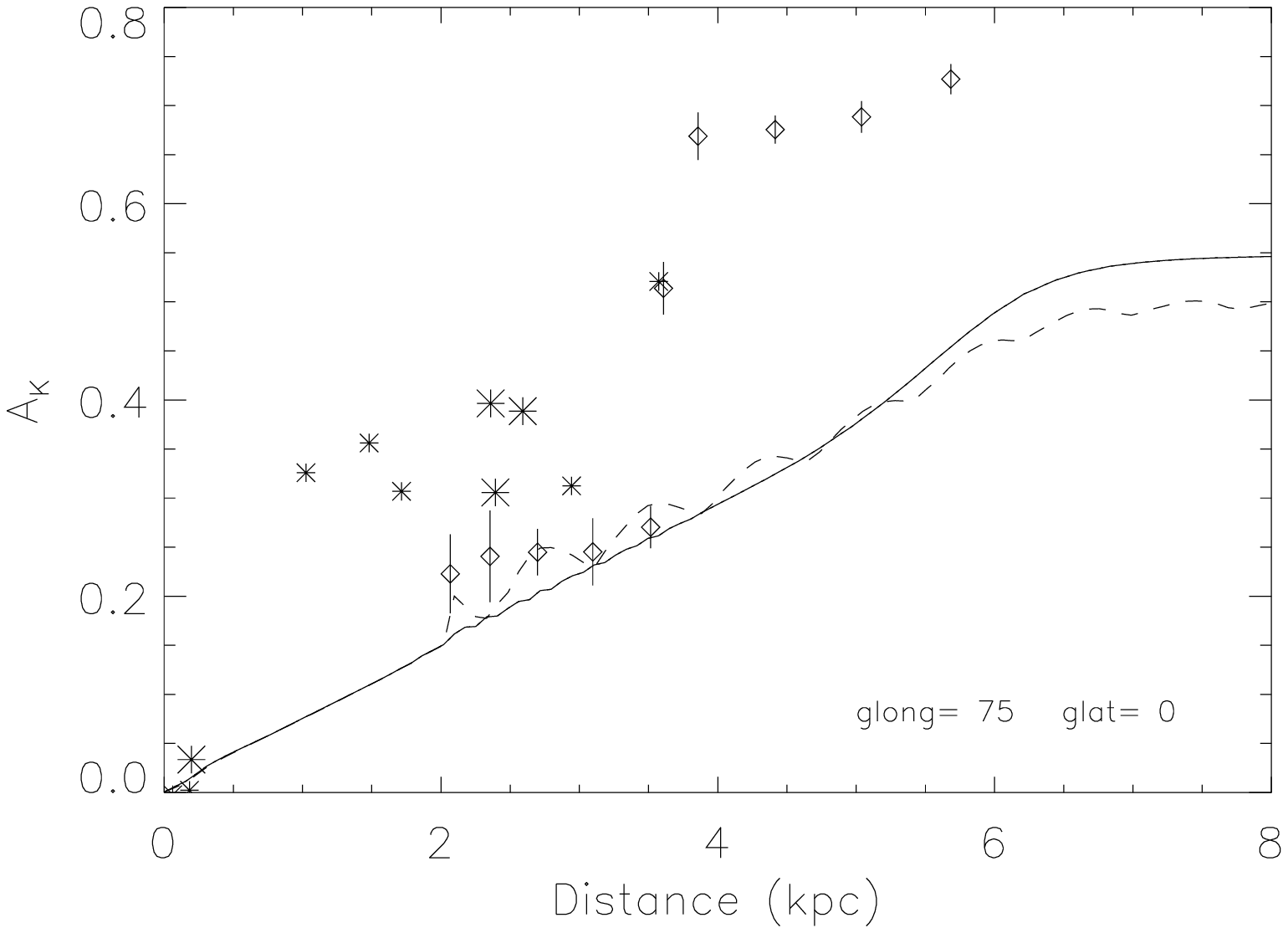}
\caption{
K band extinctions along lines-of-sight in the inner disk of the
Galaxy. Lines and symbols as in previous figure, with the addition of
a dotted line showing the extinction of the model in the case where
the disk component of the model is rescaled rather than the spiral arm
component. 
}
\label{inner}
\end{figure*}

Figs. \ref{outer} and \ref{inner} show the extinction along
selected lines-of-sight in and near the Galactic plane as given by the
extinction model and the NIR data. Also shown where available are
extinction measures from individual OB stars from the Neckel et~al.
compilation \citep{NK80, NKCat}. These extinctions are converted to
the K band using $(A_{\rm K}/A_{\rm V}) = 0.112$ as given by \citet{RL85}.
When present these give a measure of the extinction out to a
couple of kiloparsecs from the Sun, whereas the NIR red-clump data
give a measure of extinction as far as 8 kpc. In general the NIR
extinction measures are more 
reliable as each measure (distance and extinction) is based on many
stars, as described above.  
In addition the distances of the OB stars are spectro-photometric, and thus
suffer from large uncertainties due to misclassification and cosmic
variation. 

In Fig. \ref{outer} lines-of-sight with and without OB stars are
presented; in those fields containing no OB stars (left column) there
is good agreement between the model and the NIR extinction measures,
while in those fields containing many OB stars the model is
consistently giving higher extinction than indicated by the NIR data.
This can partly be understood by the effect of the OB stars
on their environment, as they will heat the ambient dust leading to
higher dust emission than is typically seen in the arms. This
contributes to the erroneous rescaling seen in the field at
$l=173^\circ$.  Also worth noting for this field, as well as that at
$l=187^\circ$, is that the two data sets give inconsistent measures of
the extinction, which must monotonically increase with distance. This
inconsistency might in part be accounted for if the extinction curve for these
lines-of-sight were nonstandard, i.e. $R_{\rm V} = 5$ rather than the
canonical value of 3.1. In this case the extinctions of the OB
stars would systematically decrease by approximately 3/5ths while
their distances would increase, giving much better
agreement with the NIR data which is insensitive to variations in the
extinction curve. In any case, the model
falls between the two empirical measures for these two fields.

For the inner disk (Fig. \ref{inner}) the deviations between the data
and the model are not as 
easily understood. Typically these deviations are of the order of
20\%, though seem to be higher in directions corresponding to tangents
of spiral arms ($l = 50^\circ$ and $30^\circ$). Also shown for some of
these lines-of-sight is the extinction curve for a model in which the disk
was rescaled rather than the nominally selected spiral arms. The performance
of this alternative rescaling is in general not significantly worse nor
better, though it seems to be better along spiral arm tangents and
worse in other directions. The exception
is in the direction $l=50^\circ$, which is much better. However, this apparent
improvement may be misleading, as the failure of the model may lie in
its assumed geometry of the Sag-Car spiral arm, which places its
tangent inside of $l=50^\circ$. The direction
$l=75^\circ$ is influenced by the local Orion-Cygnus arm, but shows
good agreement with the NIR data out to a distance of 4 kpc, at which
point the NIR data shows a large step in extinction. (The nonscaled
extinctions suffer  artifacts from interpolation errors as the single
grid containing the dust model does not have sufficiently small
gridding to resolve this feature.)  Again, for this
field and that at $l = 60^\circ$, the OB extinctions are higher than
those given by the NIR data.

\section{Summary}

We have here presented a three-dimensional Galactic extinction model,
based on the dust distribution model of DS01,
which may be used for a variety of studies involving distant
targets located in the Galactic disk or beyond. See \citet{BE03} as an
example. 

Any potential user of the extinction model should be aware of a number
of caveats: 
\begin{enumerate}
\item Absorptions to points within a few hundred parsecs of the Sun
will not be reliable as the spatial resolution of the dust model does
not allow a detailed description of dust in the vicinity of the Sun.
A more detailed local map of extinction is needed for nearby sources.

\item The dust model, even with rescaling, has finite angular
resolution. The size of the COBE pixels are approximately $0.35^\circ
\times 0.35^\circ$, with
some variation over the sky. However, any field of this size (or even
smaller) will
contain unresolved structures in the dust distribution. Thus the
run of extinction as given by the model for a given line-of-sight should
be considered as a mean extinction. 

\item The dust model does not include features associated with the
Galactic bar nor the nuclear disk, though a central ``hole'' in the
dust distribution is included. However, the parameters describing this
hole are ad hoc as the FIR data within $20^\circ$ of the Galactic
center was not modeled by DS01. In addition, the adopted spiral arm
geometry extend the arms {\bf too} far into the center of the Galaxy, leading
to arm tangents that are not evident in FIR emission in the Galactic
plane. The extinctions within $0.35R_\odot$ of the
Galactic center may suffer significant systematic errors.

\item The density of the spiral arms are not well constrained by the
COBE NIR data because the arm tangents are too distant to cause
strong extinction features in the diffuse NIR radiation;
lines-of-sight corresponding to arm tangents may have large systematic errors.

\item In directions where extragalactic
sources contribute significantly to the FIR flux the derived rescaling
factors will be erroneous. This includes the Andromeda Galaxy, M33, the LMC
and SMC. 

\item The rescaling factors in directions which contain anomalous dust
temperatures will be in error. These include the nearby star forming
regions of the Orion and rho Ophiucus complexes.
\end{enumerate}

For the purpose of validating the extinction model
we have also presented a method by which the extinction can be
empirically estimated using NIR color-magnitude diagrams. 
Comparisons made for lines-of-sight in the outer Galaxy show good
agreement between the model and the NIR data, showing differences
less than 0.05 magnitude, though directions toward
known OB associations have a high relative difference, in part due to
anamolous dust temperatures. Toward the inner Galaxy there is
less concordance, especially in directions corresponding to spiral arm
tangents. Most of our comparisons with the NIR data are in the
Galactic plane ($b = 0$). Interestingly the effect of the rescaling
factors in the Galactic plane is not as clearly advantageous as it is at
low Galactic latitudes ($|b| \approx 2 - 3^\circ$). Again, anomolous
dust temperatures associated with star forming regions may be limiting
the usefullness of the rescaling factors.

When reliable extinction data is available it is always
preferable to rely on these rather than a model. However, even when
such data is available a model may be useful to extrapolate beyond the
limits of the data, as it may be able to provide extinctions at larger
(or smaller) distances or at higher resolution than the data. The user
is also free to compute  
new rescaling factors, or to rescale the entire model, when it is in
clear conflict with data for a particular line-of-sight. 

Some of the caveats listed above suggest possible improvements
that could be made in the Galactic dust distribution model,
particularly toward the Galactic center. Work is currently underway
toward the development of a dust distribution model that also models
the FIR emission within $20^\circ$ of the Galactic center, not
originally considered in the modeling of DS01.
With the recent availability of high quality NIR data over the whole
sky, together with methods for estimating the extinction along
lines-of-sight close to the Galactic plane, it will soon be possible
to improve the extinction model by providing further observational
constraints. Such added constraints will be
particularly helpful in deriving a more reliable estimate of the
contribution of the spiral arms.

\begin{acknowledgements}
Thanks are extended to Gianpaolo Bertelli, Antonella Vallenari, Annie
Robin and Francesca Figueras for testing preliminary versions of the
extinction code, and to F. Figueras for contributing {\small FORTRAN} code. 
The first author wishes to thank Mario Lattanzi for encouragement and
moral support, while financial support was provided by the Italian Ministry of
Research (MIUR) under contract COFIN2001. 
\end{acknowledgements}

\bibliographystyle{aa}
\bibliography{H4503.bib}

\end{document}